\documentclass[12]{article}
\usepackage{fleqn}
\usepackage{graphicx}
\begin{document}
\Large
\begin{center}
{\bf Multiple Qubits as Symplectic Polar Spaces of Order Two}
\end{center}
\vspace*{.2cm}
\begin{center}
Metod Saniga$^{1}$ and Michel Planat$^{2}$

\end{center}
\vspace*{.0cm} \normalsize
\begin{center}
$^{1}$Astronomical Institute, Slovak Academy of Sciences\\
SK-05960 Tatransk\' a Lomnica, Slovak Republic\\
(msaniga@astro.sk)

\vspace*{.2cm}
and

\vspace*{.2cm} $^{2}$Institut FEMTO-ST, CNRS, D\' epartement LPMO,
32 Avenue de
l'Observatoire\\ F-25044 Besan\c con Cedex, France\\
(michel.planat@femto-st.fr)

\end{center}

\vspace*{.2cm} \noindent \hrulefill

\vspace*{.1cm} \noindent {\bf Abstract}

\noindent
It is surmised that the algebra of the Pauli operators on the Hilbert space of $N$-qubits is embodied in the geometry
of the symplectic polar space of rank $N$ and order two, $W_{2N - 1}(2)$. The operators (discarding the identity)
answer to the points of $W_{2N - 1}(2)$, their partitionings into maximally commuting subsets correspond to spreads
of the space, a maximally commuting subset has its representative in a maximal totally isotropic subspace of
$W_{2N - 1}(2)$ and, finally, ``commuting" translates into ``collinear" (or ``perpendicular").
\\ \\
{\bf MSC Codes:} 51Exx, 81R99\\
{\bf PACS Numbers:} 02.10.Ox, 02.40.Dr, 03.65.Ca\\
{\bf Keywords:} Symplectic Polar Spaces of Order Two -- N-Qubits

\vspace*{-.1cm} \noindent \hrulefill

\vspace*{.5cm}
\noindent
It is well known that a complete basis of operators in the
Hilbert space of $N$-qubits, $N \geq 2$, can be given in terms of
the Pauli operators --- tensor products of classical $2 \times 2$
Pauli matrices. Although the Hilbert space
in question is $2^{N}$-dimensional, the operators' space is of
dimension $4^{N}$. Excluding the identity matrix, the set of
$4^{N} - 1$ Pauli operators can be partitioned into $2^{N} + 1$
subsets, each comprising  $2^{N} - 1$ mutually commuting elements
[1].  The purpose of this note is to put together several important facts
supporting the view that this operators'
space can be identified with $W_{2N-1}(q=2)$, the symplectic polar
space of rank $N$ and order two.

A (finite-dimensional) classical polar space (see [2--6] for more
details) describes the geometry of a $d$-dimensional vector space
over the Galois field $GF(q)$, $V(d, q)$, carrying a non-degenerate reflexive
sesquilinear form $\sigma$. The polar space is called symplectic,
and usually denoted as $W_{d -1}(q)$,  if this form is bilinear
and alternating, i.e., if $\sigma(x, x) = 0$ for all $x \in V(d,
q)$; such a space exists only if $d=2N$, where $N$ is called its
rank. A subspace of $V(d, q)$ is called totally isotropic if
$\sigma$ vanishes identically on it. $W_{2N-1}(q)$ can then be
regarded as the space of totally isotropic subspaces of $PG(2N-1,
q)$, the ordinary $(2N - 1)$-dimensional projective space over
$GF(q)$, with respect to a symplectic form (also known as a null
polarity), with its maximal totally isotropic subspaces, also
called generators $G$, having dimension $N - 1$.  For $q=2$ this
polar space contains
\begin{equation}
|W_{2N-1}(2)| = | PG(2N-1, 2)| = 2^{2N} - 1 = 4^{N} - 1
\end{equation}
points and
\begin{equation}
|\Sigma(W_{2N-1}(2))| = (2+1)(2^{2}+1) \ldots (2^{N}+1)
\end{equation}
generators [2--4]. An important object associated with any
polar space is its {\it spread}, i.\,e., a set of generators partitioning
its points. A spread $S$ of $W_{2N-1}(q)$ is an ($N-1$)-spread of
its ambient projective space  $PG(2N-1, q)$ [4,\,5,\,7], i.\,e., a
set of ($N-1$)-dimensional subspaces of $PG(2N-1, q)$ partitioning
its points. The cardinalities of a spread and a generator of
$W_{2N-1}(2)$ thus read
\begin{equation}
|S| = 2^{N} + 1
\end{equation}
and
\begin{equation}
|G| = 2^{N} - 1,
\end{equation}
respectively [2,\,3]. Finally, it needs to be mentioned that two
distinct points of $W_{2N-1}(q)$ are called perpendicular if they
are ``isotropically" collinear, i.\,e., joined by a totally isotropic
line of $W_{2N-1}(q)$; for $q=2$ there are
\begin{equation}
\#_{\Delta} = 2^{2N-1}
\end{equation}
points that are {\it not} perpendicular to a given point of $W_{2N-1}(2)$ [2,\,3].

Now, in light of Eq.\,(1), we can identify the Pauli operators
with the points of $W_{2N-1}(2)$. If, further, we identify the
operational concept ``commuting" with the geometrical one
``perpendicular," from Eqs.\,(3) and (4) we readily see that the points lying on
generators of  $W_{2N - 1}(2)$ correspond to maximally commuting
subsets (MCSs) of operators and a spread of $W_{2N - 1}(2)$ is
nothing but a partitioning of the whole set of operators into
MCSs. From Eq.\,(2) we then infer that the operators' space
possesses $(2+1)(2^{2}+1)\ldots(2^{N}+1)$ MCSs and, finally,
Eq.\,(5) tells us that there are $2^{2N-1}$ operators that do {\it
not} commute with a given operator; the last two statements are, for $N > 2$, still conjectures
to be rigorously proven. However, the case of
two-qubits ($N=2$) is recovered in full generality [1,\,8,\,9], with the geometry behind being that of the
{\it generalized quadrangle of order two} [9] --- the simplest
nontrivial symplectic polar space.\footnote{This object can also be recognized as the projective line over
the Jordan system of the full $2 \times 2$ matrix ring with coefficients in $GF(2)$ [9].}

\vspace*{.5cm} \noindent {\bf Acknowledgements}\\
This work was partially supported by the
Science and Technology Assistance Agency under the contract $\#$
APVT--51--012704, the VEGA project $\#$ 2/6070/26 (both from
Slovak Republic) and the trans-national ECO-NET project $\#$
12651NJ ``Geometries Over Finite Rings and the Properties of
Mutually Unbiased Bases" (France).

\end{document}